\documentclass[journal]{IEEEtran}
\ifCLASSINFOpdf
  \usepackage[pdftex]{graphicx}
  \usepackage{amsmath}
  \usepackage{amsfonts}
  \usepackage{xcolor}
  \usepackage{caption}
  \usepackage{subcaption}
  \usepackage{stfloats}
  \usepackage{cite}
\else
\fi
\begin{document}
%
% paper title
% Titles are generally capitalized except for words such as a, an, and, as,
% at, but, by, for, in, nor, of, on, or, the, to and up, which are usually
% not capitalized unless they are the first or last word of the title.
% Linebreaks \\ can be used within to get better formatting as desired.
% Do not put math or special symbols in the title.
\title{Assessing BIOMASS Radiometric Stability using Permanent Scatterers}

\author{Marco~Manzoni, %~\IEEEmembership{Member,~IEEE,}
        %Francesco~Banda, 
        Francesco~Salvaterra,
        Claudio Maria Prati
        and Stefano~Tebaldini
        %~\IEEEmembership{Senior~Member,~IEEE}% <-this % stops a space
        \thanks{This work has been submitted to the IEEE for possible
publication. Copyright may be transferred without notice,
after which this version may no longer be accessible.}%
        }

% The paper headers
\markboth{Journal of \LaTeX\ Class Files,~Vol.~14, No.~8, August~2015}%
{Shell \MakeLowercase{\textit{et al.}}: Bare Demo of IEEEtran.cls for IEEE Journals}

% make the title area
\maketitle

\begin{abstract}
Monitoring the relative radiometric stability of the European Space Agency’s BIOMASS P-band synthetic aperture radar (SAR) requires separating instrumental and processing effects from temporal changes in the scene backscatter. This study investigates the use of permanent scatterers (PSs) as opportunistic references for estimating residual radiometric gains without dedicated calibration targets. The PS-based calibration (PSCal) framework combines amplitude-based scatterer selection with iterative maximum-likelihood gain estimation. Simulations assess detection and estimation performance under the limited number of acquisitions available in BIOMASS stacks, highlighting the importance of stringent selection thresholds and the dependence of estimation accuracy on scatterer abundance and signal-to-noise ratio. The approach is applied to datasets over Nairobi, Kenya, and Córdoba, Argentina. Over Nairobi, PSCal identifies an approximately 1 dB radiometric anomaly subsequently traced to a defect in radio-frequency interference mitigation, while the remaining acquisitions exhibit estimated gain variations generally within 0.1 dB. Over Córdoba, twelve acquisitions spanning two mission phases separated by approximately eight months yield residual gain variations on the order of 0.1 dB despite pronounced changes in the scene backscatter and acquisition geometry. Results were observed to be consistent with measurements from the BIOMASS Calibration Transponder (BCT) installed in New Norcia (Australia). These results demonstrate the potential of carefully selected P-band PSs to complement dedicated calibration infrastructure for relative radiometric monitoring and the detection of processing anomalies.
\end{abstract}


\IEEEpeerreviewmaketitle

\section{Introduction}
\IEEEPARstart{R}{adiometric} calibration is essential for the quantitative exploitation of spaceborne synthetic aperture radar (SAR) data \cite{freeman_sar_1992, schwerdt_independent_2017, klenk_independent_2026, geudtner_sentinel-1_2013}. This requirement is particularly important for low-frequency P-band systems, such as the ESA BIOMASS mission \cite{quegan_european_2019}, whose long wavelength provides sensitivity to forest structure and above-ground biomass \cite{soja_mapping_2021,ho_tong_minh_capabilities_2015}. Since radiometric measurements contribute directly to above-ground biomass estimation \cite{carreiras_estimating_2013, gama_eucalyptus_2010}, radiometric errors can propagate into the retrieved biomass estimates. Radio-frequency interference (RFI) can further compromise image radiometry, as can the techniques used to mitigate it. These techniques typically suppress interference in the frequency domain, either before or after image focusing \cite{leanza_leo_2019}. Unless properly implemented, such processing can also alter the amplitude of the desired SAR signal, introducing radiometric distortions into the resulting images.\\
Conventional calibration strategies typically rely on a combination of internal calibration devices and external reference targets, such as corner reflectors or transponders \cite{wang_performance_2021}. While these approaches provide high accuracy during the commissioning phase, their long-term availability and global coverage are inherently limited. At P-band, in fact, implementing conventional external calibration infrastructure becomes particularly challenging. The long wavelength implies that corner reflectors must be physically very large to ensure sufficient radar cross section, making their design, deployment, and maintenance both complex and costly \cite{algafsh_measurements_2017}. Similarly, active transponders operating at P-band require specialized hardware, high power, and careful design, which significantly limits their practical use \cite{barbary_biomass_2023}. As a consequence, unlike higher-frequency missions, there is currently no global network of well-characterized P-band calibration targets available \cite{gisinger_-depth_2021}. This lack of dedicated infrastructure severely constrains the applicability of traditional calibration strategies, especially beyond the commissioning phase. Moreover, the sparse spatial distribution of any potential calibrators would prevent consistent monitoring of radiometric stability at a global scale. In this context, relying solely on external references becomes unsustainable for long-term mission operations. Therefore, alternative approaches that do not depend on artificial targets are required. An alternative and complementary solution is to exploit naturally occurring stable targets in the scene, commonly referred to as permanent scatterers (PSs, \cite{ferretti_permanent_2001}), whose amplitude stability can be leveraged to monitor temporal variations in the system gain \cite{daria_sar_2010}.

PS-based calibration techniques have been extensively investigated for C- and X-band SAR missions, for which long time series comprising several tens of acquisitions are often available \cite{guccione_low-frequency_2018}. Under these conditions, a large number of independent observations (typically more than 30) improves estimation accuracy and can allow performance to approach asymptotic bounds. In contrast, BIOMASS provides a limited number of repeat-pass acquisitions over a given area, with typical tomographic stacks comprising only $N_i = 7$ images. Combining two stacks covering the same area increases the number of observations and may therefore improve calibration accuracy. However, differences in mean incidence angle and limited footprint overlap restrict the set of PSs that can be consistently identified across both stacks. Suitable PSs must lie within the overlapping area and exhibit a sufficiently stable backscattering response over the incidence angles spanned by the two stacks.

The rest of the paper is organized as follows: Section \ref{sec:simulation} presents the PS detection and radiometric gain estimation framework and evaluates its performance through simulations. Section \ref{sec:real_data} applies the method to BIOMASS datasets acquired over Nairobi and Córdoba and provides a qualitative comparison with BIOMASS Calibration Transponder measurements. Section \ref{sec:conclusions} summarizes the main findings. The appendices discuss the bias in the estimated dispersion index and its relationship with the signal-to-noise ratio.

\section{PS detection, parameter estimation and simulated performances}
\label{sec:simulation}
In this section, we briefly recap the theory underlying PSs and the estimation of radiometric gains using them. A more detailed analysis is carried out in \cite{ferretti_permanent_2001, daria_sar_2010, guccione_low-frequency_2018, monti_guarnieri_long-term_2013, guccione_stable_2012}.
\begin{figure}[t]
\centering
\includegraphics[width=\columnwidth]{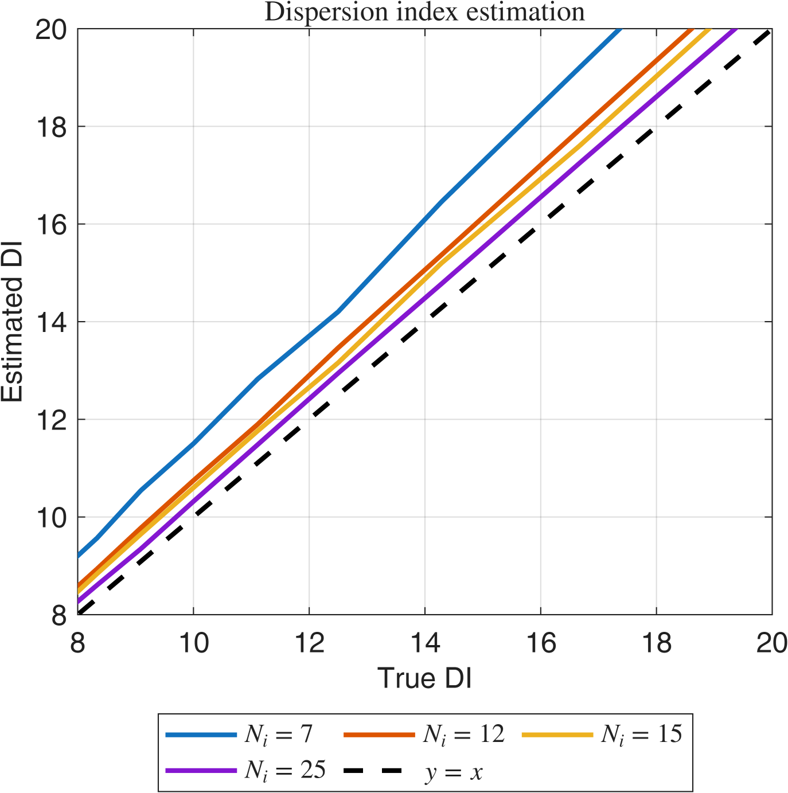}
\caption{True vs estimated dispersion index for varying number of images. The estimated DI is biased towards higher values when the number of images in the stack is low. The bias becomes worse the higher the true DI.}
\label{fig:bias_dispersion_index}
\end{figure} 
\subsection{Signal model}
Let us consider $N_\mathrm{i}$ images and $N_\mathrm{p}$ persistent scatterers (PSs) in the scene. The complex observation of the $p$th target in the $n$th image is modeled as \cite{daria_sar_2010}:
\begin{equation}
    y_n^p = a_n \left(b_p e^{j\phi_n^p} + w_n^p\right),
    \label{eq:model}
\end{equation}
where $a_n>0$ is the amplitude gain of the $n$th image, $b_p \geq 0$ is the amplitude of the $p$th PS, $\phi_n^p$ is its phase, and $w_n^p$ represents additive clutter and thermal noise before multiplication by the image gain. The image gains may vary between acquisitions but are assumed to be spatially uniform over the analyzed scene. Conversely, the PS amplitudes may differ between targets but are assumed to remain constant across acquisitions. This complementary dependence on the image and target indices allows the image gains and PS amplitudes to be separated up to a common multiplicative factor. We resolve this ambiguity by setting $a_1=1$, so that the estimated gains are relative to the first image and the PS amplitudes are expressed on its amplitude scale.
The observation model in \eqref{eq:model} can be written in vector form as
\begin{equation}
    \mathbf{y} = \mathbf{x} + \mathbf{w}_a,
\end{equation}
where $\mathbf{y} \in \mathbb{C}^{N_\mathrm{t}}$, with $N_\mathrm{t}=N_\mathrm{p}N_\mathrm{i}$, contains the observations stacked target by target, with the image index varying fastest:
\begin{equation}
    \mathbf{y}
    =
    \left[
    y_1^1 \ \cdots \ y_{N_\mathrm{i}}^1 
    y_1^2 \ \cdots \ y_{N_\mathrm{i}}^2 
    \cdots 
    y_1^{N_\mathrm{p}} \ \cdots \ y_{N_\mathrm{i}}^{N_\mathrm{p}}
    \right]^{\mathsf{T}}.
\end{equation}
Using the same ordering for all stacked quantities, we define
\begin{align}
    \mathbf{x}
    &= (\mathbf{b} \otimes \mathbf{a}) \circ \boldsymbol{\phi},\\
    \mathbf{b}
    &= [b_1 \ \cdots \ b_{N_\mathrm{p}}]^{\mathsf{T}},\\
    \mathbf{a}
    &= [1 \ a_2 \ \cdots \ a_{N_\mathrm{i}}]^{\mathsf{T}},\\
    [\boldsymbol{\phi}]_{(p-1)N_\mathrm{i}+n}
    &= e^{j\phi_n^p},\\
    [\mathbf{w}_a]_{(p-1)N_\mathrm{i}+n}
    &= a_n w_n^p.
\end{align}
Here, $\otimes$ denotes the Kronecker product, $\circ$ denotes the Hadamard product, and the superscript $\mathsf{T}$ denotes transposition. Our objective is to estimate the relative image gains $\mathbf{a}$, together with the nuisance parameters $\mathbf{b}$ and $\boldsymbol{\phi}$ and the parameters describing the noise power.
\begin{figure}[t]
\centering
\includegraphics[width=\columnwidth]{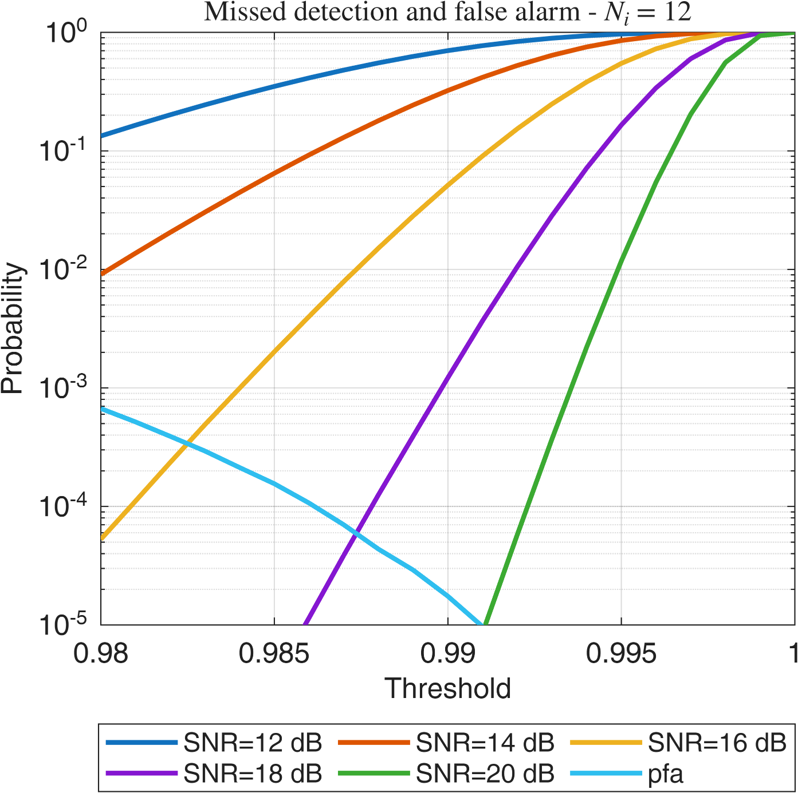}
\caption{Probability of false alarm (pfa) and missed detection (pmd) for the incoherent detector at varying thresholds and SNRs. The number of images is fixed to 7.}
\label{fig:pfa_pmd_incoherent}
\end{figure} 
\subsection{Incoherent detection}
The first step in the processing pipeline is PS detection. Incoherent detection discards phase information and relies on image amplitudes. In particular, \cite{daria_sar_2010} showed that the dispersion index (DI) provides a figure of merit equivalent to that of the generalized likelihood ratio test (GLRT):
\begin{equation}
\label{eq:dispersion_index}
    \mathrm{DI} = \frac{\mu}{\sigma}
\end{equation}
where $\mu$ and $\sigma$ are the temporal mean and standard deviation of the amplitude of each pixel, respectively. A target exhibiting high reflectivity and small temporal fluctuations is therefore a suitable PS candidate.
A transformation of DI was proposed in \cite{daria_sar_2010} to obtain a figure of merit ranging from 0 to 1. Here, we refer to this quantity as the normalized dispersion index (NDI):
\begin{equation}
    \mathrm{NDI}= \frac{1}{\sqrt{1+\frac{1}{DI^2}}}
\end{equation}
which approaches 1 as DI tends to infinity and 0 as DI tends to zero.
When only a few images are available, the DI estimate is biased. This bias is directly proportional to the true dispersion index and inversely proportional to the number of images in the stack:
\begin{equation}
\mathbb{E}[\hat{DI}] \approx \frac{\mu}{\sigma} + \underbrace{\frac{\mu}{2N\sigma}}_{\mathrm{bias}}
\label{eq:bias_eq}
\end{equation}
\begin{figure}[t]
\centering
\includegraphics[width=\columnwidth]{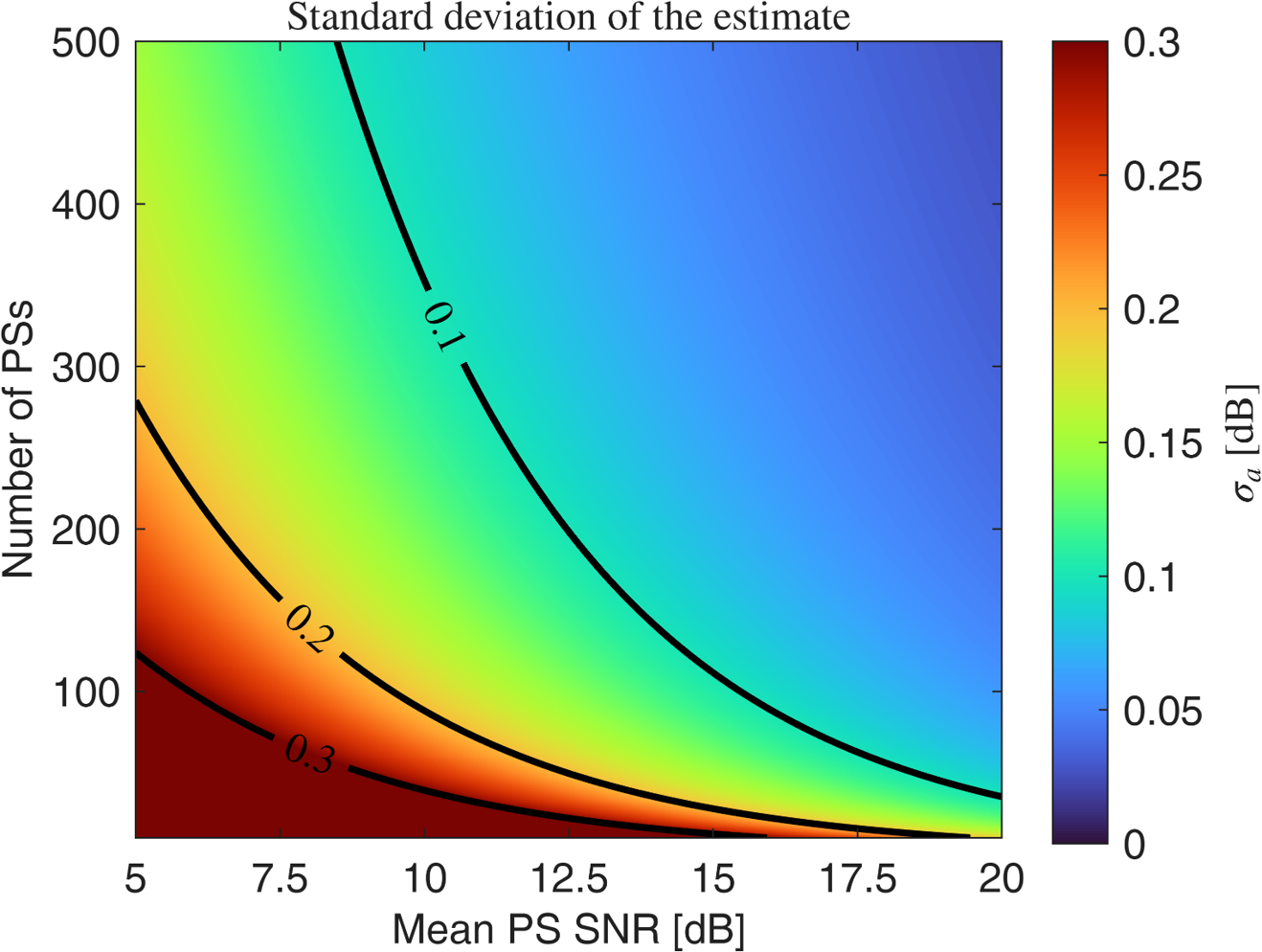}
\caption{Cramér-Rao Bound (CRB) for the estimate of the radiometric gains for varying number of PSs and mean SNR.}
\label{fig:CRB}
\end{figure} 
The derivation of \eqref{eq:bias_eq} is provided in Appendix \ref{app:bias}. Figure \ref{fig:bias_dispersion_index} presents simulation results for $N_i=7,12,15,25$, showing that the estimated dispersion index increasingly departs from its nominal value as the stack size decreases, particularly at high nominal DI values. The limited number of acquisitions, combined with the high temporal stability achievable with P-band measurements, motivates the use of a high DI threshold. For example, an estimated DI threshold of 16 for a stack of 7 images leads to the detection of PSs with a true DI of approximately 14.\\
Coherent PS detection is also possible, but it requires phase estimation. Assuming a structured phase model, such as a linear model, the phase history can be estimated and compensated before applying coherent detection, as described in \cite{daria_sar_2010}. Here, however, we favor incoherent detection because it avoids phase estimation, which can be challenging when only a few acquisitions are available. Accordingly, only incoherent detection is considered in the following.\\
Following \cite{guccione_low-frequency_2018}, we also apply an amplitude threshold. For low-frequency missions such as BIOMASS, distributed scatterers can exhibit high temporal stability as well. Furthermore, the DI bias associated with short stacks motivates an additional selection criterion to reduce false alarms. In the following analysis, we select the brightest $2\%$ of pixels in the scene and compute DI only for this subset.\\
An alternative approach is to require the backscatter to exceed the instrument's noise-equivalent sigma zero (NESZ) by at least 10 dB. For BIOMASS, a NESZ of $-27$ dB yields a backscatter threshold of $\sigma_0>-17$ dB.
\subsection{Detection performances}
We simulated a set of PSs across a stack of 12 images, with SNR values ranging from 12 to 20 dB, and evaluated the probability of missed detection ($P_\mathrm{MD}$) and the probability of false alarm ($P_\mathrm{FA}$) for different NDI thresholds. The stack size was chosen to match the number of single-look complex (SLC) images in the real dataset analyzed in the final section of this contribution. Figure \ref{fig:pfa_pmd_incoherent} presents the results for the incoherent detector, which relies exclusively on the amplitude information of each PS candidate. Increasing the threshold reduces the probability of incorrectly identifying clutter as a PS, at the cost of a higher probability of missed detection. For example, an NDI threshold of $0.99$, corresponding to a DI threshold of approximately $7.02$ through $\mathrm{DI}=\mathrm{NDI}/\sqrt{1-\mathrm{NDI}^2}$, yields a false-alarm probability of approximately $10^{-5}$. At this threshold, the probability of missed detection is approximately $80\%$ for a PS with an SNR of 12 dB and decreases to $0.1\%$ at an SNR of 18 dB.
\begin{figure}[t]
\centering
\includegraphics[width=\columnwidth]{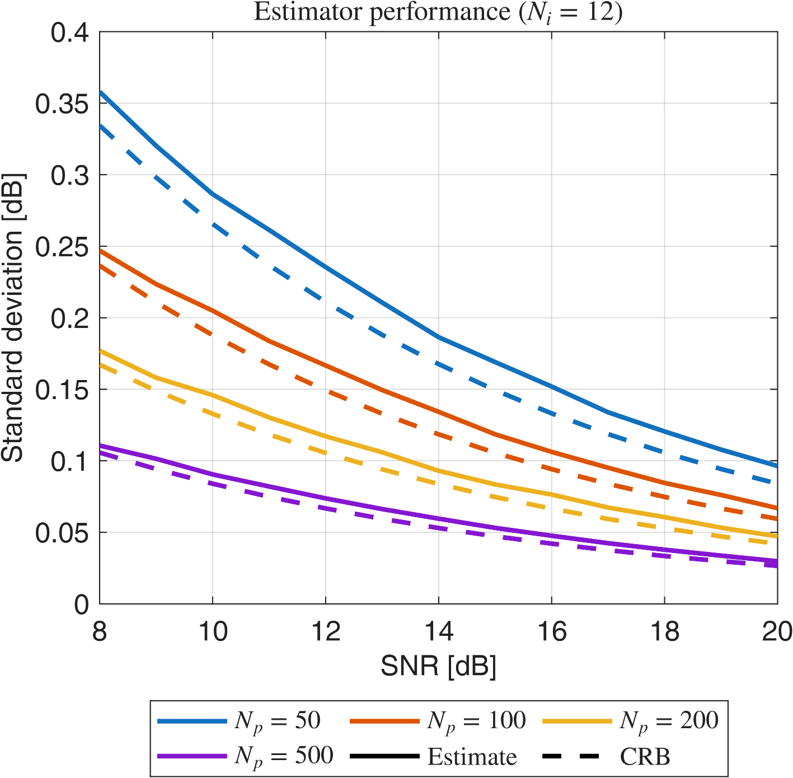}
\caption{Simulated performances in terms of standard deviation on the estimate of the residual differential gains. As an example, an accuracy of 0.1 dB can be achieved using 500 PSs with an average SNR of 10 dB, or 200 PSs with an average SNR of 14 dB.}
\label{fig:performances}
\end{figure} 
\subsection{Incoherent estimation}
The log-likelihood of the observation can be expressed as follows (omitting a constant term that is irrelevant for the maximization):
\begin{equation}
\label{eq:loglikelihood}
 \mathcal{L}(\mathbf{y}|\mathbf{x},\mathbf{C}) = -(\mathbf{y}-\mathbf{x})^{\mathrm{H}} \mathbf{C}^{-1}(\mathbf{y}-\mathbf{x})-\log{|\mathbf{C}|}   
\end{equation}
If the PS phases are completely unknown, we can replace in \eqref{eq:loglikelihood} the maximum likelihood solution:
\begin{equation}
    \hat{\phi}_n^p = -\angle{y_n^p}
\end{equation}
leading to the expression of the log-likelihood in \eqref{eq:incoherent_loglikelihood}.

\begin{figure*}[!b]
\begin{equation}
    \mathcal{L}=2\sum_{n=1}^{N_\mathrm{i}}\frac{1}{a_n}\sum_{p=1}^{N_\mathrm{p}}\frac{b_p|y_n^p|}{v_p}-
    \sum_{n=1}^{N_\mathrm{i}}\sum_{p=1}^{N_\mathrm{p}}\frac{|y_n^p|^2}{a_n^2v_p}-
    N_\mathrm{i}\sum_{p=1}^{N_\mathrm{p}}\left(\frac{b_p^2}{v_p}-\log v_p\right)-2N_\mathrm{p}\sum_{n=1}^{N_\mathrm{i}}\log a_n
    \label{eq:incoherent_loglikelihood}
\end{equation}
\end{figure*}
The estimation procedure is performed iteratively. Starting from an initial set of gains, these values are used to estimate the PS amplitudes and the noise variances. The resulting estimates are then employed to update the gains, and the process is repeated until convergence.
At each iteration, the PS amplitudes are estimated via maximum likelihood by setting to zero the derivative of the log-likelihood in \eqref{eq:incoherent_loglikelihood} with respect to $\mathbf{b}$,
\begin{equation}
\frac{\partial \mathcal{L}(\mathbf{y}|\mathbf{a}^{k-1},\mathbf{b},\mathbf{v})}{\partial \mathbf{b}}=0
\end{equation}
which yields
\begin{equation}
\hat{b}^k_p = \frac{1}{N_\mathrm{i}}\sum_{n=1}^{N_\mathrm{i}}\frac{|y_n^p|}{a^{k-1}_n}.
\end{equation}
This estimate corresponds to the normalized temporal mean of each PS.
The algorithm then estimates the noise power associated with each PS by nullifying the derivative of the log-likelihood with respect to $\mathbf{v}$
\begin{equation}
\frac{\partial \mathcal{L}(\mathbf{y}|\mathbf{a}^{k-1},\mathbf{b}^k,\mathbf{v})}{\partial \mathbf{v}}=0,
\end{equation}
leading to
\begin{equation}
\hat{v}^k_p = \frac{1}{N_\mathrm{i}}\sum_{n=1}^{N_\mathrm{i}}\frac{|y_n^p|^2}{(a_n^{k-1})^2} - (\hat{b}^k_p)^2
\end{equation}
Finally, the gains are updated by setting to zero the derivative of the log-likelihood with respect to $\mathbf{a}$
\begin{equation}
\frac{\partial \mathcal{L}(\mathbf{y}|\mathbf{a},\mathbf{b}^k,\mathbf{v}^k)}{\partial \mathbf{a}}=0,
\end{equation}
which results in a second-order equation. The physically meaningful estimate of the gains is obtained by selecting the largest solution of
\begin{equation}
N_\mathrm{p}a_n^2 + \sum_{p=1}^{N_\mathrm{p}}\frac{b_p|y_n^p|}{v_p}a_n - \sum_{p=1}^{N_\mathrm{p}}\frac{|y_n^p|^2}{v_p}=0.
\end{equation}
The estimated gains are relative quantities and can therefore be recovered only up to an unknown constant multiplicative factor. Resolving this ambiguity requires an additional constraint or an external reference, such as assuming that the residual gains average to unity in linear scale (or zero in the logarithmic domain) over a sufficiently large number of acquisitions, or using an externally calibrated reflector. Nevertheless, as demonstrated in the following section, the estimated gains provide a valuable basis for assessing radiometric stability and may serve as an early warning indicator of hardware or software malfunctions.
\subsection{Estimation performances}
Before evaluating the estimation performance in terms of the standard deviation of the gain estimates, in \cite{daria_sar_2010} a derivation of the Cramér–Rao Bound (CRB) of the estimate of the residual gains is presented, resulting in 
\begin{equation}
    \sigma_\mathrm{CRB}^2 = \frac{1}{2N_p\overline{\mathrm{SNR}}}
\end{equation}
where $N_p$ is the number of PSs and $\overline{\mathrm{SNR}}$ is the average Signal-To-Noise Ratio (SNR) of all the detected PSs. 
The CRB for the amplitude gain $a>0$ is converted to the decibel domain by considering $g(a)=20\log_{10}a$. The CRB transformation rule gives
\begin{equation}
    \mathrm{CRB}_{g}
    = \left(\frac{20}{a\ln 10}\right)^2 \sigma_\mathrm{CRB}^2
\end{equation}
Leading to the corresponding standard-deviation bound as $\approx8.6\,\sigma_\mathrm{CRB}/a$ in decibels. For small relative uncertainties, this also approximates the standard deviation of the transformed estimate through first-order error propagation. An equivalent first-order expression is
\begin{equation}
    20\log_{10}\left(1+\frac{\sigma_{\mathrm{CRB}}}{a}\right)
    \simeq \frac{20}{\ln 10}\frac{\sigma_\mathrm{CRB}}{a}
\end{equation}
This metric already provides a preliminary figure of merit for the expected uncertainty. Figure \ref{fig:CRB} shows how the CRB varies with both the mean SNR and the number of detected PSs, highlighting a clear trade-off between these quantities. In particular, accurate estimation can still be achieved at low SNR if a sufficiently large number of PSs is available, whereas high SNR enables reliable estimates even with fewer PSs. The black curves in Fig.\ref{fig:CRB} represent standard deviations of 0.1, 0.2, and 0.3 dB. For example, reaching a 0.1 dB standard deviation requires roughly 150 PSs at an average SNR of 15 dB, or about 400 PSs when the average SNR is 10 dB.\\
To properly assess the performance of the coherent and incoherent estimators across different SNRs and numbers of PSs, a simulation campaign was conducted. In particular, $N_p = 50, 100, 200, 500$ PSs were generated with SNRs ranging from 8 dB to 20 dB over a stack of $N_i = 12$ images. The incoherent estimator was then applied to estimate the residual radiometric gains. Using the ground truth gains imposed on each image, the estimation error was computed as
\begin{align}
e_n^{\mathrm{dB}} &= \hat{a}_n^{\mathrm{dB}} - a_n^{\mathrm{dB}}, \\
\sigma_a &= \sqrt{\frac{1}{N_i}\sum_n \left(e_n^{\mathrm{dB}} - \overline{e}^{\mathrm{dB}}\right)^2 } .
\end{align}
Figure \ref{fig:performances} show the estimator performance for $N_p = 50, 100, 200$, respectively. A common behavior observed across all experiments is that the standard deviation of the estimates approaches the CRB at high SNR. As expected, higher SNR values and/or a larger number of PSs result in improved estimation accuracy, ranging from approximately 0.35 dB for $N_p = 50$ at SNR $= 8$ dB to values close to 0.05 dB for $N_p = 500$ at SNR $= 20$ dB.
\begin{table*}[t]
\centering
\caption{Nairobi dataset}
\label{tab:nairobi_dataset}
\resizebox{\textwidth}{!}{%
\begin{tabular}{ll}
\hline
\textbf{Date} & \textbf{Filename} \\
\hline
2025-05-25 & \texttt{\detokenize{BIO_S1_STA__1S_20250525T032423_20250525T032445_C_G__M__C__T___F001_01_DBT3Q4}} \\
2025-05-28 & \texttt{\detokenize{BIO_S1_STA__1S_20250528T032422_20250528T032444_C_G__M__C__T___F001_01_DBT3R1}} \\
2025-05-31 & \texttt{\detokenize{BIO_S1_STA__1S_20250531T032422_20250531T032444_C_G__M__C__T___F001_01_DBT3RW}} \\
2025-06-03 & \texttt{\detokenize{BIO_S1_STA__1S_20250603T032422_20250603T032444_C_G__M__C__T___F001_01_DBT3ST}} \\
2025-06-06 & \texttt{\detokenize{BIO_S1_STA__1S_20250606T032422_20250606T032445_C_G__M__C__T___F001_01_DBT3TQ}} \\
2025-06-09 & \texttt{\detokenize{BIO_S1_STA__1S_20250609T032423_20250609T032445_C_G__M__C__T___F001_01_DBT3UN}} \\
2025-06-12 & \texttt{\detokenize{BIO_S1_STA__1S_20250612T032422_20250612T032444_C_G__M__C__T___F001_01_DBT3VL}} \\
2025-06-15 & \texttt{\detokenize{BIO_S1_STA__1S_20250615T032422_20250615T032444_C_G__M__C__T___F001_01_DBT3WJ}} \\
2025-06-18 & \texttt{\detokenize{BIO_S1_STA__1S_20250618T032421_20250618T032444_C_G__M__C__T___F001_01_DBT3XG}} \\
2025-06-21 & \texttt{\detokenize{BIO_S1_STA__1S_20250621T032424_20250621T032446_C_G__M__C__T___F001_01_DBT3YC}} \\
\hline
\end{tabular}%
}
\end{table*}
\begin{figure*}[t]
    \centering
    \includegraphics[width=\textwidth]{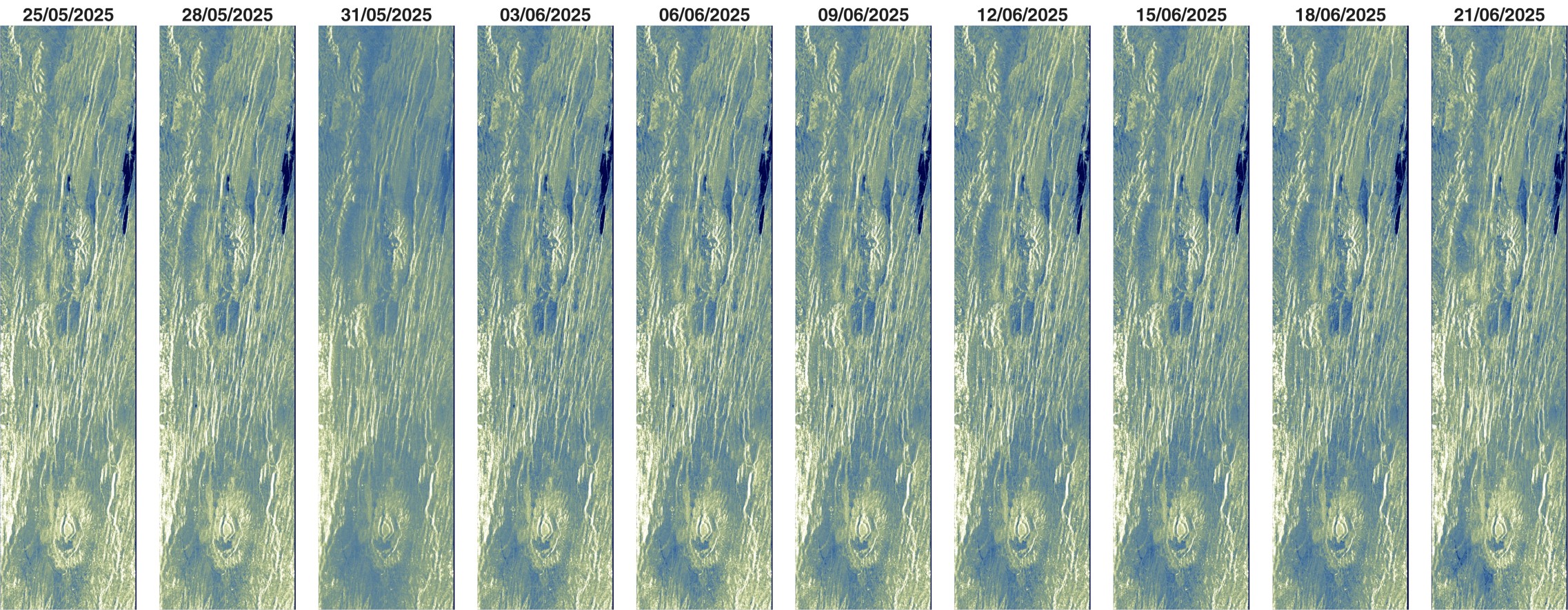}
    \caption{Multilooked amplitude images of the Nairobi stack in VV polarization. The third acquisition (31 May 2025) exhibits a radiometric anomaly. Subsequent analysis traced the anomaly to the RFI mitigation strategy applied during processing, which distorted the signal spectrum and consequently altered the acquisition’s radiometry.}
    \label{fig:Nairobi_scene}
\end{figure*}
\section{Real data analysis}
\label{sec:real_data}
\begin{figure}[t]
\centering
\includegraphics[width=\columnwidth]{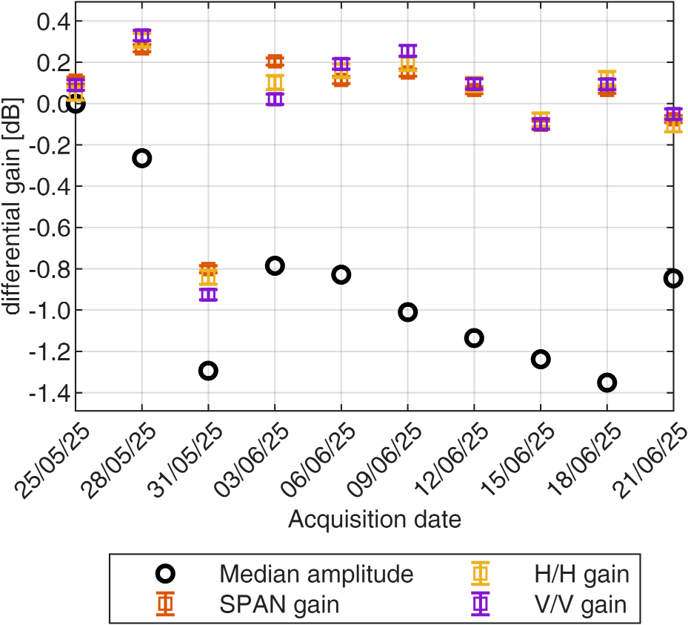}
\caption{Residual radiometric gains of the Nairobi dataset with a detected anomaly in the third image. In red, yellow, and purple, the SPAN, HH, and VV polarizations, respectively. In black circles, the median amplitude of the entire scene. }
\label{fig:Nairobi_gains}
\end{figure} 
To assess PS availability across different scene types and validate the algorithm’s ability to reliably estimate residual gains, we applied the complete procedure to several datasets acquired during the In-Orbit Commissioning Phase (IOCP) and the subsequent tomographic phase. This section presents results from selected test cases, illustrating the method’s potential for identifying processing-chain anomalies and assessing the instrument’s long-term radiometric stability.\\
First, the Level-1A (L1A) images are downloaded from the Multi-Mission Algorithm and Analysis Platform (MAAP). These data are already calibrated, and polarimetric corrections have been applied. The SLC stack is then processed through a two-step coregistration procedure combining geometric alignment with local cross-correlation to ensure accurate registration across all acquisitions. Next, the dispersion index is computed according to Eq.~\eqref{eq:dispersion_index} for each polarimetric channel as well as for the span, defined as the image whose pixel amplitude is given by $\sqrt{|VV|^2 + |VH|^2 + |HV|^2 + |HH|^2}$.
\subsection{Nairobi - Kenya}
During the IOCP, several test sites were analyzed using PSCal, including an area surrounding Nairobi, Kenya. The dataset comprises ten acquisitions separated by three-day intervals, spanning 27 days (Table~\ref{tab:nairobi_dataset}). The nominal geometrical baselines were below 1~km, consistent with the acquisition configuration used during the early stages of the IOCP. All images were radiometrically calibrated using the lookup tables (LUTs) provided in the product metadata, such that multilooking the squared magnitude of the calibrated samples yields an estimate of the backscatter coefficient $\sigma_0$.\\
Visual inspection of Fig.~\ref{fig:Nairobi_scene} reveals a pronounced radiometric anomaly in the third acquisition, dated 31 May 2025. An initial assessment can be obtained by comparing image histograms or median amplitudes across the stack. The black markers in Fig.~\ref{fig:Nairobi_gains} show the median amplitude relative to the first acquisition and indicate a substantial decrease in the third image. However, scene-wide statistics alone cannot distinguish an instrumental or processing anomaly from changes associated with environmental conditions, such as rainfall and soil moisture, or propagation effects, such as ionospheric disturbances.\\
PS candidates were identified using the incoherent detector, with a minimum dispersion index of 16, corresponding to an SNR of approximately 21~dB. The minimum backscatter was set for a minimum $\sigma_0$ of 7~dB. The latter threshold is well above the instrument's nominal NESZ requirement of $-27$~dB. These criteria yielded 735 PS candidates in SPAN, 184 in HH, and 334 in VV, whereas no candidates were detected in the cross-polarized channels. The density of PS is 0.5 PS/$\mathrm{km}^2$ in the SPAN, 0.21 PS/$\mathrm{km}^2$ in the VV and 0.12 PS/$\mathrm{km}^2$ in the HH. \\
The larger number of PS candidates identified using the span may partly result from a reduction in relative temporal dispersion when summing the powers of different polarimetric channels. When their fluctuations are not perfectly positively correlated, the summed power can exhibit a lower relative dispersion than the individual contributions. This reduction becomes more pronounced when fluctuations in different channels partially compensate for one another. Consequently, more pixels may satisfy the dispersion threshold, even in the absence of a single dominant, temporally stable scatterer.\\
The corresponding residual radiometric gains, estimated using PSCal, are shown in Fig.~\ref{fig:Nairobi_gains}.\\
The SPAN, HH, and VV estimates exhibit consistent behavior across the stack and identify a decrease of approximately 1~dB between the first and third acquisitions. Conversely, the decreasing trend in median amplitude from the fourth acquisition onward is absent from the PSCal estimates. This difference is consistent with a change in scene backscatter rather than a common radiometric gain variation. By restricting the estimation to appropriately selected stable scatterers, PSCal reduces sensitivity to scene-wide backscatter changes that affect histogram- and median-based assessments. Excluding the anomalous acquisition, the estimated gain variations are generally within 0.1~dB, indicating high relative radiometric stability over the observation period.\\
Subsequent investigation established that the third acquisition was the only image in the stack processed with radio-frequency interference (RFI) mitigation enabled. A defect in the initial implementation of this processing step distorted the signal spectrum and consequently altered the image radiometry. This case illustrates the value of PSCal as an early indicator of radiometric anomalies and its ability to help distinguish common gain variations from changes in scene backscatter.
\begin{figure}[t]
\centering
\includegraphics[width=\columnwidth]{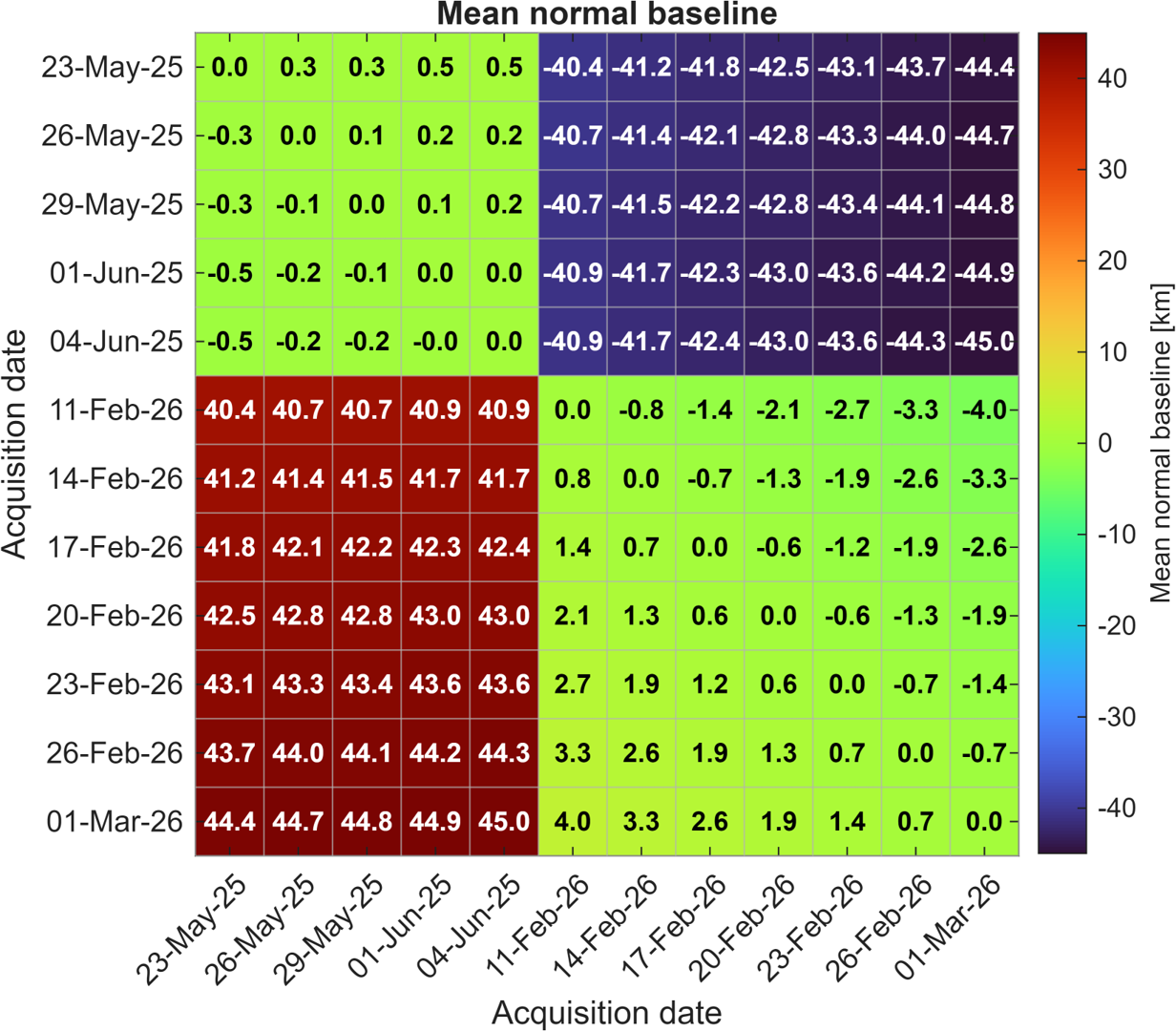}
\caption{Normal baselines in the Argentinian dataset.}
\label{fig:baselines_corodba}
\end{figure} 

\begin{table*}[t]
    \centering
    \caption{Argentina dataset}
    \label{tab:argentina_dataset}
    \resizebox{\textwidth}{!}{%
    \begin{tabular}{ll}
        \hline
        \textbf{Date} & \textbf{Filename} \\
        \hline
        2025-05-23 & \texttt{\detokenize{BIO_S1_STA__1S_20250523T102053_20250523T102144_C_G__M__C__T___F001_01_DVEZ0X}} \\
        2025-05-26 & \texttt{\detokenize{BIO_S1_STA__1S_20250526T102052_20250526T102143_C_G__M__C__T___F001_01_DVEZ2G}} \\
        2025-05-29 & \texttt{\detokenize{BIO_S1_STA__1S_20250529T102052_20250529T102143_C_G__M__C__T___F001_01_DVEZ3X}} \\
        2025-06-01 & \texttt{\detokenize{BIO_S1_STA__1S_20250601T102051_20250601T102143_C_G__M__C__T___F001_01_DVEZ5E}} \\
        2025-06-04 & \texttt{\detokenize{BIO_S1_STA__1S_20250604T102052_20250604T102143_C_G__M__C__T___F001_01_DVEZ6V}} \\
        2026-02-11 & \texttt{\detokenize{BIO_S1_STA__1S_20260211T102239_20260211T102330_T_G01_M02_C01_T021_F001_01_DVEZ8G}} \\
        2026-02-14 & \texttt{\detokenize{BIO_S1_STA__1S_20260214T102241_20260214T102332_T_G01_M02_C02_T021_F001_01_DVEZ9O}} \\
        2026-02-17 & \texttt{\detokenize{BIO_S1_STA__1S_20260217T102243_20260217T102334_T_G01_M02_C03_T021_F001_01_DVEZB8}} \\
        2026-02-20 & \texttt{\detokenize{BIO_S1_STA__1S_20260220T102244_20260220T102336_T_G01_M02_C04_T021_F001_01_DVEZCV}} \\
        2026-02-23 & \texttt{\detokenize{BIO_S1_STA__1S_20260223T102246_20260223T102337_T_G01_M02_C05_T021_F001_01_DVEZEJ}} \\
        2026-02-26 & \texttt{\detokenize{BIO_S1_STA__1S_20260226T102248_20260226T102339_T_G01_M02_C06_T021_F001_01_DVEZG5}} \\
        2026-03-01 & \texttt{\detokenize{BIO_S1_STA__1S_20260301T102250_20260301T102341_T_G01_M02_C07_T021_F001_01_DVEZHR}} \\
        \hline
    \end{tabular}%
    }
\end{table*}
\subsection{Córdoba - Argentina}
\begin{figure}[t]
\centering
\includegraphics[width=\columnwidth]{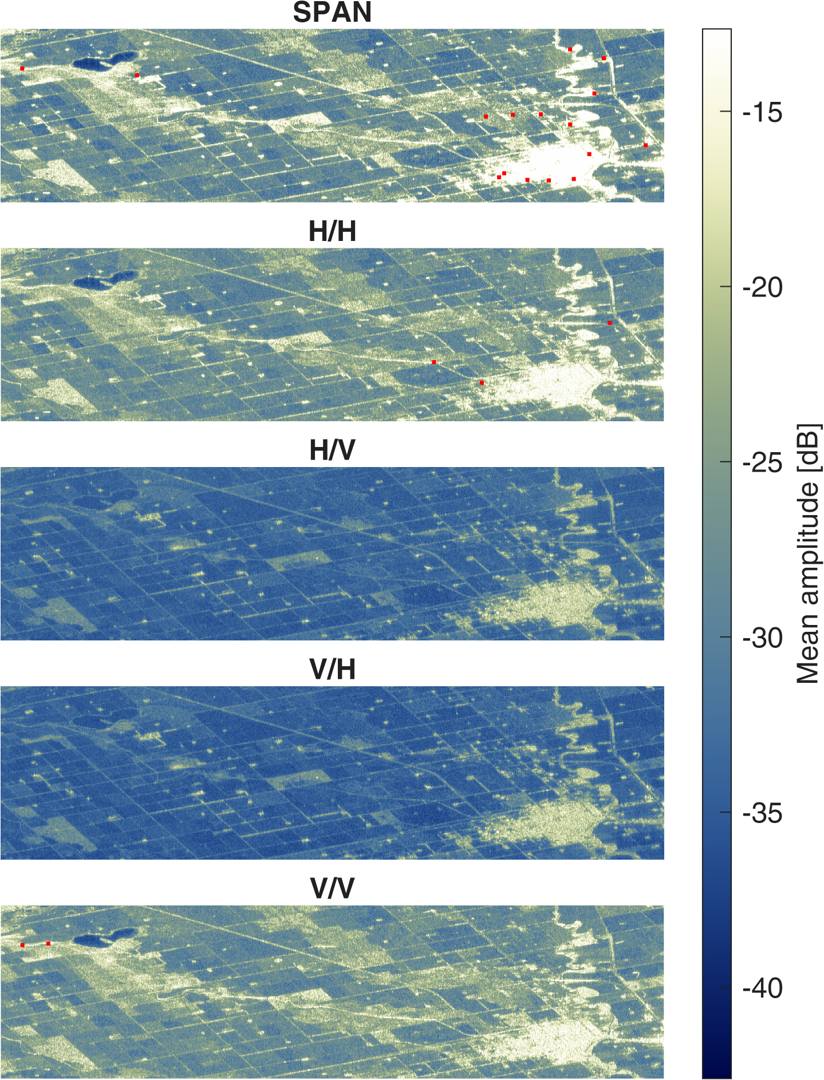}
\caption{Zoom over the Argentinian site. PSs are generally man-made objects. Some of them are present only in the HH, only in the VV or only in the span.}
\label{fig:corodba_zoom}
\end{figure}

\begin{figure}[t]
\centering
\includegraphics[width=\columnwidth]{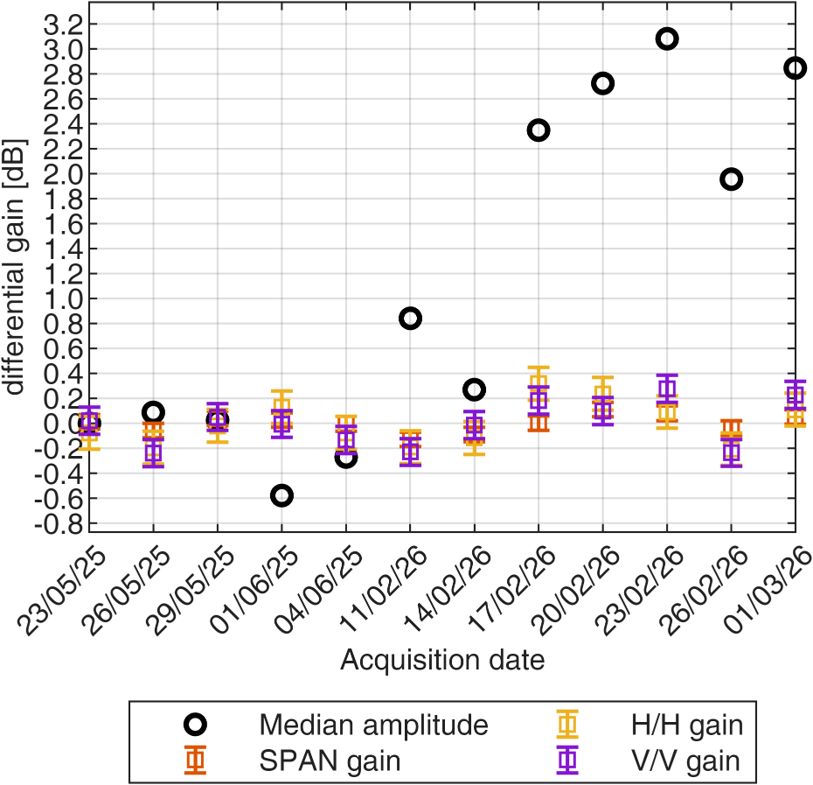}
\caption{Residual radiometric gains estimated by PSCal for the Argentina dataset using SPAN (red), HH (yellow), and VV (purple). Black circles indicate the median amplitude over the entire scene. With appropriate PS selection, the estimated gains do not follow the temporal trend of the scene median, suggesting that PSCal distinguishes instrumental radiometric variations from changes in scene backscatter.}
\label{fig:cordoba_gains}
\end{figure}

The second dataset covers Argentina and combines two sub-stacks acquired during different mission phases. The first comprises five IOCP acquisitions, while the second consists of seven acquisitions collected during the tomographic phase. As reported in Table~\ref{tab:argentina_dataset}, both sub-stacks were acquired at three-day intervals, spanning 12 and 18 days, respectively, with an approximately eight-month gap between them. This extended observation period allows assessment of the instrument's long-term relative radiometric stability.\\
Combining the two sub-stacks introduces three main challenges. First, the longer observation period may reduce the number of scatterers that remain stable compared with a single, regularly sampled stack covering a shorter period. Second, the different acquisition geometries result in a substantial inter-stack baseline and a corresponding change in incidence angle. Third, the reduced overlap between the acquisition footprints limits the area available for PS detection. Consequently, jointly selected PSs must exhibit stable backscatter both over time and across the sampled viewing geometries. This requirement favors targets with limited angular sensitivity, although it does not imply isotropic scattering.\\
Figure~\ref{fig:baselines_corodba} shows the mean normal baseline for each of the twelve acquisitions. The first five acquisitions have approximately zero relative baseline, whereas the seven tomographic acquisitions span normal baselines of up to approximately 4~km within their sub-stack. The two sub-stacks are separated by a normal baseline of approximately 40~km.\\
\begin{figure*}[t]
    \centering
    \includegraphics[width=\textwidth]{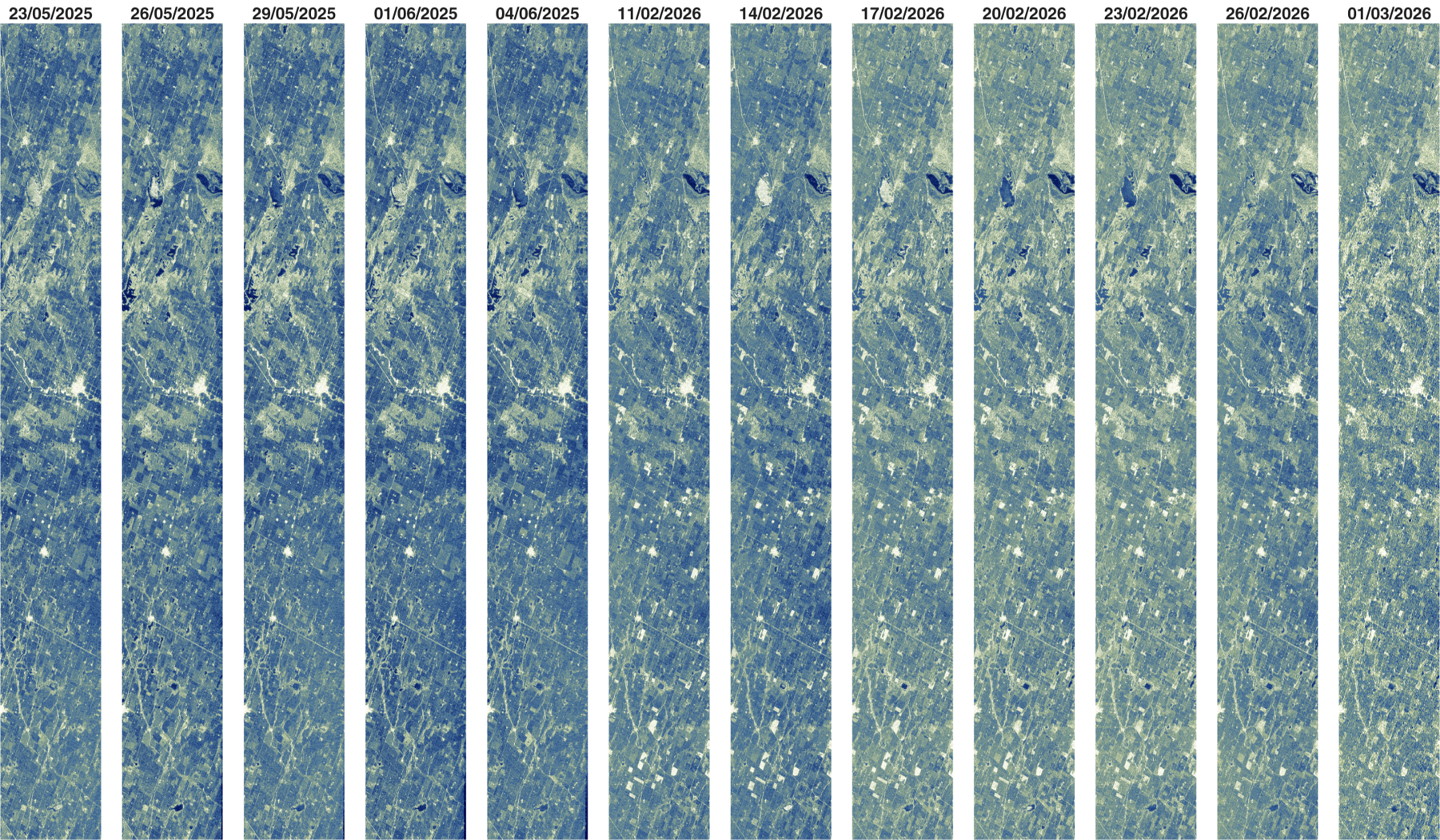}
    \caption{Single-look amplitude images of the Argentina stack. Starting from the 17\textsuperscript{th} of February 2026, the image appears brighter due to a radical change in the backscatter.}
    \label{fig:cordoba_scene}
\end{figure*}
Visual inspection reveals a change in the temporal evolution of the average scene backscatter beginning on February 17 (Fig.~\ref{fig:cordoba_scene}). The images become brighter, potentially reflecting changes in the scene, the instrument response, or the processing pipeline. This increase is also evident from the black markers in Fig.~\ref{fig:cordoba_gains}, which represent the median amplitude of each image normalized to that of the first acquisition in the stack.\\
The larger number of acquisitions compared with the previous case study supports a more reliable estimation of the DI, allowing a less restrictive selection threshold. PS candidates were identified using the incoherent detector, with a minimum dispersion index of 12.5, corresponding to an SNR of approximately 19~dB, and a minimum $\sigma_0$ of $-17$~dB. The latter threshold remains well above the instrument's nominal NESZ requirement of $-27$~dB. These criteria yielded 105 PS candidates in SPAN, 21 in HH, and 29 in VV, while no candidates were detected in the cross-polarized channels. The density of PSs in this case is much lower than in the previous case due to scene composition. The density of PSs is 0.1 PS/$\mathrm{km}^2$ in the SPAN and roughly 0.03 PS/$\mathrm{km}^2$ in the VV and HH. In Figure \ref{fig:corodba_zoom} a zoomed amplitude image with the detection (red dots) of the PSs found. Most of them are man-made objects. The residual radiometric gains estimated using PSCal are shown in Fig.~\ref{fig:cordoba_gains}.\\
The gain estimates exhibit consistent oscillations across the available polarization channels, with variations on the order of 0.1~dB. Moreover, their temporal evolution does not follow the changes in median scene amplitude. These observations suggest that the pronounced scene brightening is not primarily driven by a common instrumental gain variation. In general, the measured amplitude of each pixel depends on both the target response and the radiometric response of the acquisition and processing chain. If scene-induced variations dominate the instrumental fluctuations, selecting sufficiently stable scatterers can suppress the influence of scene variability and enable the estimation of residual radiometric gains. This interpretation assumes that the selected scatterers remain stable throughout the observation period.
\subsection{Qualitative BIOMASS Calibration Transponder (BCT) comparison}
The residual radiometric gains estimated by PSCal over Argentina are compared with the BCT measurements in the HH and VV channels during the overlapping observation period in February 2026. For each polarization, a constant offset, computed as the difference between the means of the two full time series, is subtracted from the BCT measurements to account for their different reference levels.\\
The comparison is necessarily qualitative, as the two approaches employ different measurement and estimation techniques, observe different sites, and sample different acquisition dates. Neither point-by-point agreement nor identical temporal trends should therefore be expected. Moreover, the offset correction enables a comparison of relative variations but precludes an assessment of agreement in absolute radiometric level. An additional limitation is that the BCT-derived gains do not account for potential fluctuations in the transponder response, which may lead to either an overestimation or an underestimation of the instrument's radiometric variability.\\
Despite these limitations, both datasets exhibit fluctuations on the order of a few tenths of a decibel. Although the PSCal estimates show larger excursions over the selected interval, their magnitude remains comparable to that of the BCT measurements. This agreement in scale supports the plausibility of the residual radiometric variations retrieved by PSCal, while providing a qualitative consistency check rather than a direct validation of the individual gain estimates.
\begin{figure}[t]
\centering
\includegraphics[width=\columnwidth]{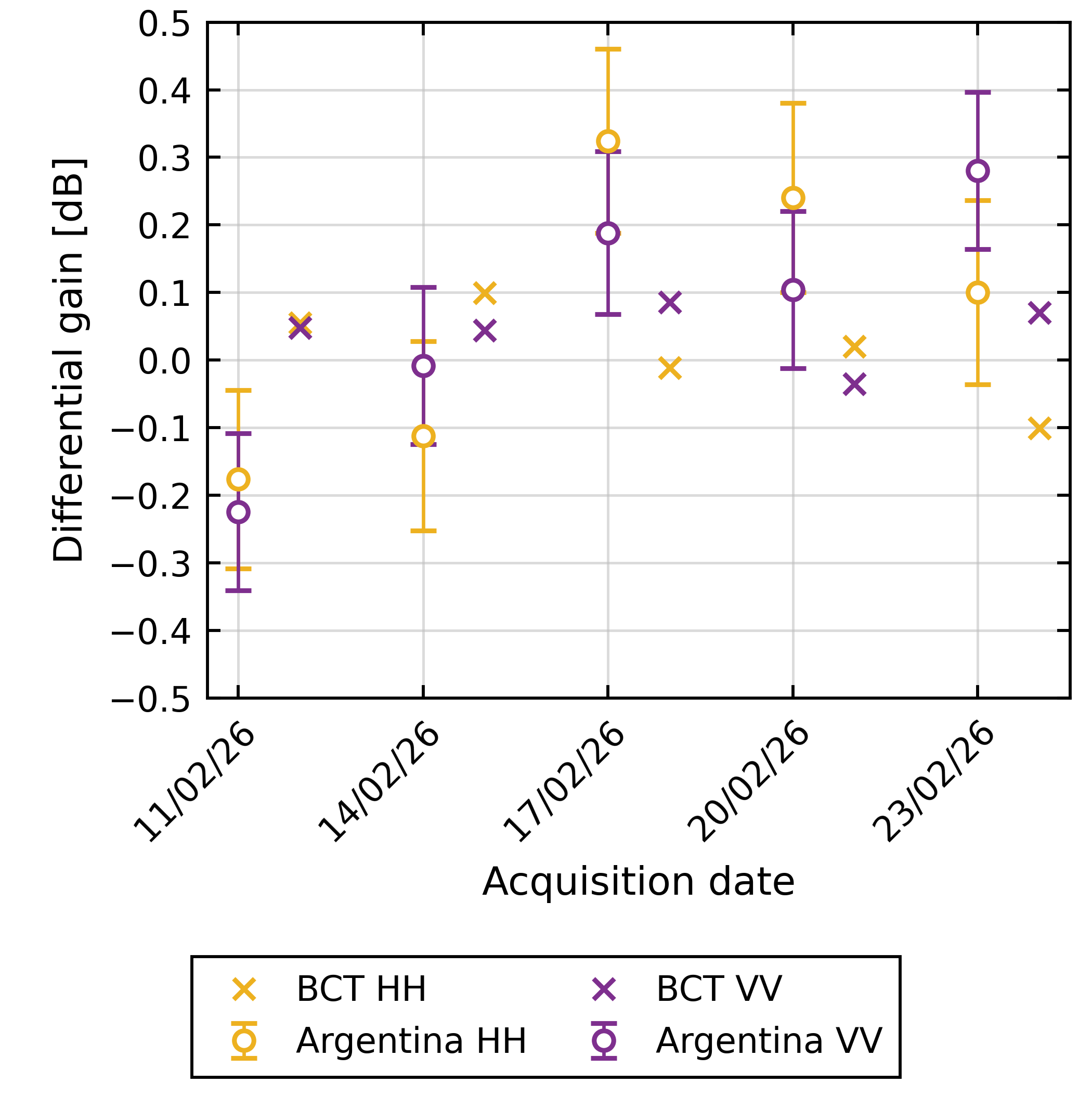}
\caption{Qualitative comparison between the BCT and the PSCal-derived gains.}
\label{fig:BCT_Argentina_IEEE_errorbars}
\end{figure} 
\section{Conclusion}
\label{sec:conclusions}
This study investigated the use of permanent scatterers to assess the relative radiometric stability of BIOMASS P-band SAR instrument. The analysis combined amplitude-based PS detection, iterative estimation of residual radiometric gains, and simulations representative of the limited stack sizes available during the mission. The results highlight PS selection as a critical step: with few acquisitions, uncertainty in the estimated dispersion index complicates the identification of stable targets, requiring stringent selection criteria. Simulations further show that gain estimation accuracy improves with both the number of available PSs and their signal-to-noise ratio, approaching the theoretical performance bound at high SNR.\\
The real-data experiments demonstrate the value of PSCal for identifying radiometric anomalies and assessing relative gain variations in the presence of changing scene backscatter. Over Nairobi, the method detected an approximately 1~dB anomaly subsequently attributed to a defect in the RFI mitigation processing, while the remaining acquisitions exhibited estimated gain variations generally within 0.1~dB. Over Córdoba, the joint analysis of two sub-stacks separated by approximately eight months yielded residual gain variations on the order of 0.1~dB, despite substantial differences in acquisition geometry and pronounced changes in median scene amplitude. The qualitative comparison with BIOMASS Calibration Transponder measurements showed fluctuations of comparable magnitude, supporting the plausibility of the PSCal estimates without providing a direct validation.\\
These findings support the use of PSCal as a complementary tool for monitoring the relative radiometric response of the instrument and processing chain. Its interpretation remains conditional on the temporal and angular stability of the selected scatterers, and the estimated gains do not provide an absolute radiometric reference. Future work should establish an operational criterion for selecting detection thresholds that balances scatterer reliability and estimation uncertainty, extend the assessment to additional sites and acquisition configurations, and pursue closer temporal alignment with independent calibration measurements.

\appendix[Bias on the estimate of the dispersion index]
\label{app:bias}
Let $A_1, \dots, A_N$ be i.i.d. random variables with mean $\mu$ and variance $\sigma^2$. The dispersion index is defined as
\begin{align}
DI = \frac{\mu}{\sigma},
\end{align}
and its empirical estimate is
\begin{align}
\hat{DI} = \frac{\hat{\mu}}{\hat{\sigma}},
\end{align}
where
\begin{align}
\hat{\mu} = \frac{1}{N} \sum_{i=1}^N A_i, 
\qquad
\hat{\sigma}^2 = \frac{1}{N} \sum_{i=1}^N (A_i - \hat{\mu})^2.
\end{align}

Define the function $g(m,v) = m / \sqrt{v}$, so that $DI = g(\mu,\sigma^2)$. A first-order Taylor expansion of $g(\hat{\mu}, \hat{\sigma}^2)$ around $(\mu,\sigma^2)$ yields
\begin{align}
\hat{DI} \approx \frac{\mu}{\sigma}
+ \frac{\hat{\mu} - \mu}{\sigma}
- \frac{\mu}{2\sigma^3} (\hat{\sigma}^2 - \sigma^2).
\end{align}

Taking expectations and using $\mathbb{E}[\hat{\mu}] = \mu$ and $\mathbb{E}[\hat{\sigma}^2] = \frac{N-1}{N}\sigma^2$, we obtain
\begin{align}
\mathbb{E}[\hat{DI}] \approx \frac{\mu}{\sigma} + \frac{\mu}{2N\sigma}.
\end{align}

Thus, the estimator is biased with
\begin{align}
\mathrm{Bias}(\hat{DI}) \approx \frac{\mu}{2N\sigma}.
\end{align}

\appendix[Relationship Between Incoherent Dispersion Index and SNR in SAR Images]
Consider a complex SAR pixel modeled as
\begin{equation}
Z = A + N,
\end{equation}
where \(A>0\) is the deterministic signal amplitude and \(N\) is circular complex Gaussian noise,
\begin{equation}
N = N_I + jN_Q,
\end{equation}
with
\begin{equation}
N_I,N_Q \sim \mathcal{N}(0,\sigma_n^2),
\end{equation}
independent and identically distributed. The total mean noise power is therefore
\begin{equation}
\mathbb{E}\left[|N|^2\right]
=
\mathbb{E}\left[N_I^2+N_Q^2\right]
=
2\sigma_n^2.
\end{equation}
Defining the SNR as the ratio between signal power and total complex noise power gives
\begin{equation}
\mathrm{SNR}
=
\frac{A^2}{\mathbb{E}[|N|^2]}
=
\frac{A^2}{2\sigma_n^2}.
\end{equation}
Now consider the incoherent amplitude image,
\begin{equation}
R = |Z|
=
\sqrt{(A+N_I)^2+N_Q^2}.
\end{equation}
The incoherent dispersion index is defined as
\begin{equation}
D
=
\frac{\mu_R}{\sigma_R}
=
\frac{\mathbb{E}[R]}
{\sqrt{\operatorname{Var}(R)}}.
\end{equation}
In the high-SNR regime,
\begin{equation}
A \gg \sigma_n,
\end{equation}

the amplitude \(R\) can be expanded a tthe first order around \(A\):
\begin{equation}
R \simeq A+N_I.
\end{equation}
Hence,
\begin{equation}
\mathbb{E}[R]
\simeq A,
\end{equation}
and
\begin{equation}
\operatorname{Var}(R)
\simeq
\operatorname{Var}(N_I)
=
\sigma_n^2.
\end{equation}
Thus,
\begin{equation}
D
=
\frac{\mu_R}{\sigma_R}
\simeq
\frac{A}{\sigma_n}.
\end{equation}
Comparing this expression with the one for the SNR, we obtain:
\begin{equation}
\mathrm{SNR}_{\mathrm{dB}}
\simeq
20\log_{10}
\left(
\frac{\mu_R}{\sigma_R}
\right)
-
3~\mathrm{dB}.
\end{equation}

\section*{Acknowledgment}
The authors would like to thank the entire BIOMASS In-Orbit Commissioning Phase (IOCP) team and the Aresys team for their support in preparing the data. Special thanks are extended to André Silva and the DLR team for conducting the BCT analyses.
\ifCLASSOPTIONcaptionsoff
  \newpage
\fi
\bibliographystyle{ieeetr}  % oppure plain, unsrt, alpha, ecc.
\bibliography{references}

\begin{IEEEbiography}[{%
    \includegraphics[
        width=1in,
        height=1.25in,
        clip,
        keepaspectratio
    ]{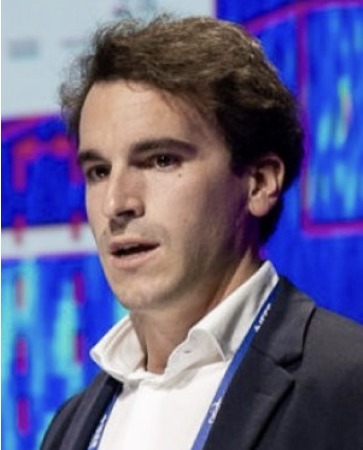}%
}]{Marco Manzoni}
(Member, IEEE) was born in Lecco, Italy, in 1994. He is a Tenure-Track Assistant Professor at Politecnico di Milano, specializing in radar remote sensing and signal processing. He received his M.Sc. in Telecommunications Engineering (2018) and Ph.D. in Information Technology (2022), both cum laude, from the same institution.
His research covers SAR imaging across spaceborne, airborne, UAV, and automotive platforms, including MIMO-SAR, motion compensation, autofocusing, interferometry, tomography, radiometric calibration, and distributed multistatic SAR systems. He contributed to the ESA Sentinel-1 C and D commissioning phases and the BIOMASS commissioning phase. His honors include Best Paper Awards at EuSAR 2024 and the 2022 Mediterranean Microwave Symposium, the 2023 PIERS Young Scientist Award, and the 2022 Dimitris N. Chorafas Award. He teaches the courses of Radar Imaging for the Ms.C. in telecommunication engineering and Advanced Radar Systems for the Ph.D. program in Information Technology at Politecnico di Milano.
\end{IEEEbiography}

\begin{IEEEbiography}[{%
    \includegraphics[
        width=1in,
        height=1.25in,
        clip,
        keepaspectratio
    ]{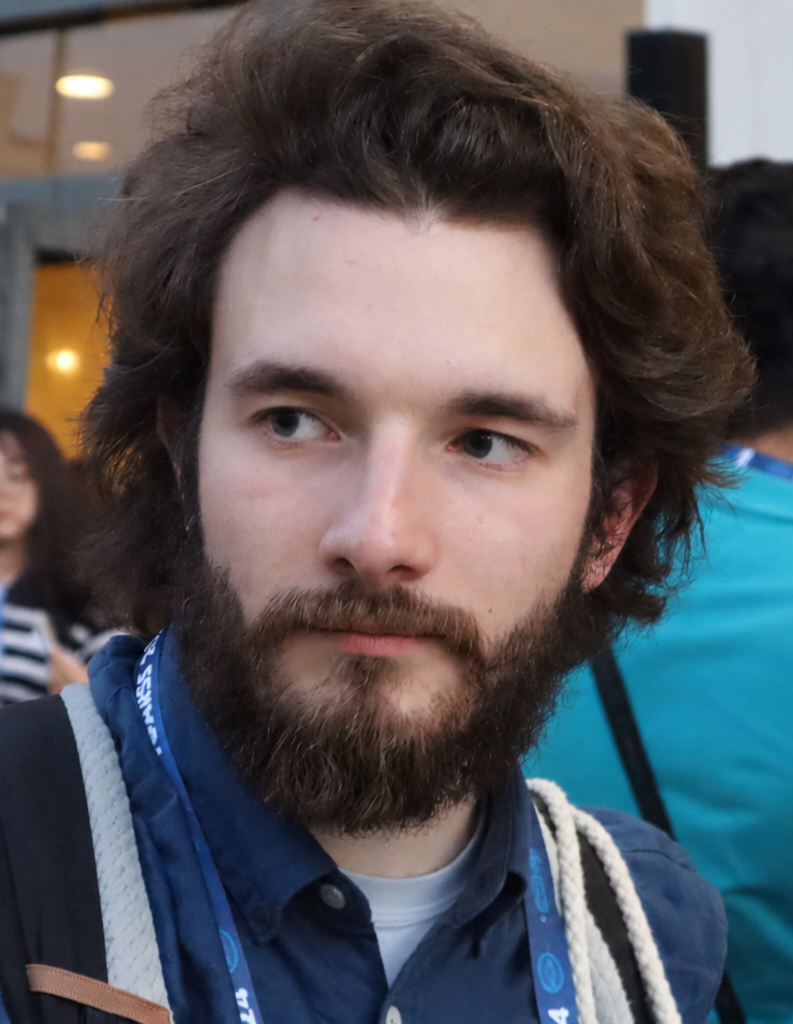}%
}]{Francesco Salvaterra}
was born in Brescia, Italy. He received the B.Sc. degree in Aerospace Engineering in 2021 and the M.Sc. degree in Space Engineering in 2023, both from Politecnico di Milano, Milan, Italy. He is currently pursuing the Ph.D. degree in Telecommunication Engineering at the same institution. His research focuses on synthetic aperture radar (SAR) interferometry and tomography, with particular emphasis on low-frequency SAR calibration, ionospheric disturbances correction, high-resolution spectral methods, and multistatic SAR performances estimation. His work includes the development of advanced processing algorithms for the European Space Agency (ESA) BIOMASS mission. From 2024 to 2025, he was a Visiting Ph.D. Student at ISAE-SUPAERO, Toulouse, France. He received the IEEE GRSS Symposium Prize Paper Award at IGARSS 2024 for his research on bistatic SAR correlation tomography.
\end{IEEEbiography}

\begin{IEEEbiography}[{%
    \includegraphics[
        width=1in,
        height=1.25in,
        clip,
        keepaspectratio
    ]{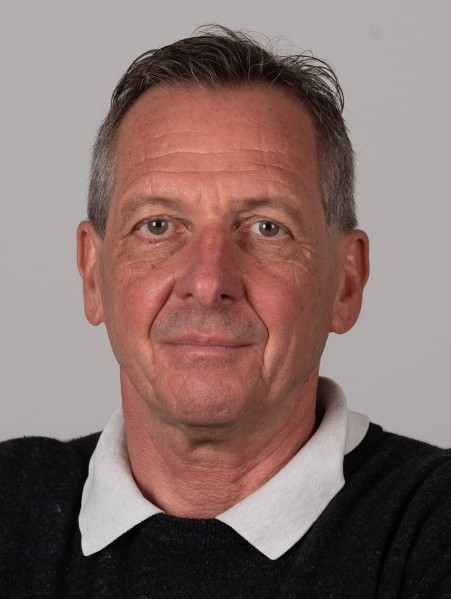}%
}]{Claudio Maria Prati}
is currently a Full Professor of telecommunications with the Electronic Department, Politecnico of Milan (POLIMI), Milan, Italy. He holds five patents in the field of SAR and SAS data processing. He has published more than 150 papers on SAR and SAS data processing and interferometry. He is the Key Scientist in several studies
by European Space Agency (ESA), European Union
(EU), the Italian National Research Council (CNR),
the Italian Space Agency (ASI), and ENI-AGIP.
He is the Co-Founder of Tele-Rilevamento Europa
(T.R.E)—a spin-off company of POLIMI that has recently become T.R.E
Altamira, a CLS French Group Company.
Dr. Prati received three prizes from the IEEE Geoscience and Remote
Sensing Society (IGARSS ’89, IGARSS ’99, and best TGARSS paper 2016).
He was the Chair of the Telecommunications Study Council at POLIMI.
\end{IEEEbiography}

% if you will not have a photo at all:
% \begin{IEEEbiographynophoto}{Francesco Salvaterra}
% Biography text here.
% \end{IEEEbiographynophoto}

% insert where needed to balance the two columns on the last page with
% biographies
%\newpage

\begin{IEEEbiography}[{%
    \includegraphics[
        width=1in,
        height=1.25in,
        clip,
        keepaspectratio
    ]{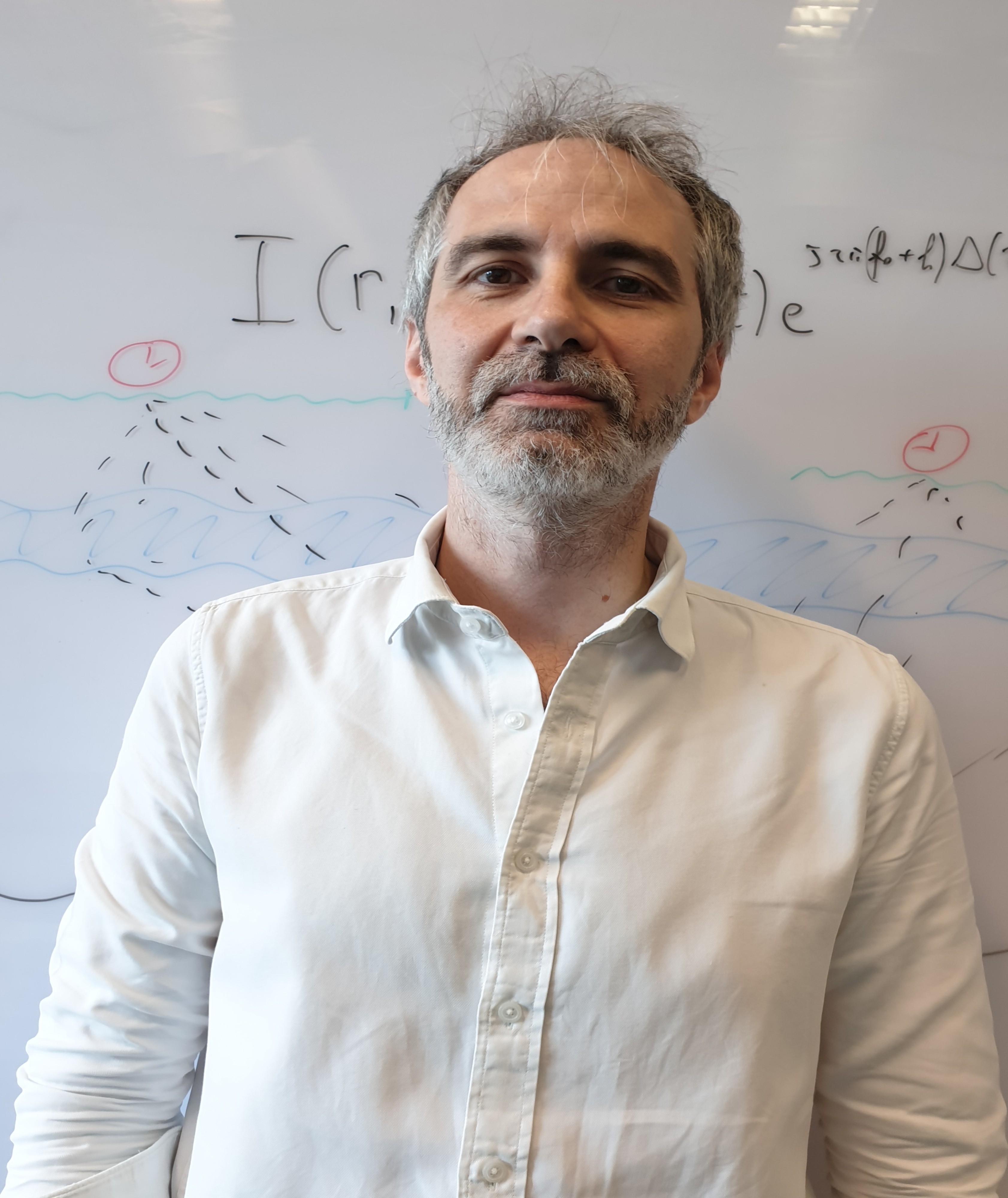}%
}]{Stefano Tebaldini}
Stefano Tebaldini received his M.Sc. in Telecommunication Engineering (2005) and Ph.D. (2009) from Politecnico di Milano, where he became Associate Professor in 2018 and Full Professor in 2026. His research focuses on Synthetic Aperture Radar (SAR) for Earth observation, spanning system design, signal processing, electromagnetic scattering, and geophysical retrieval. His scientific contributions include 3-D imaging of forests, snow, glaciers, and ice sheets, with methods for vegetation characterization, probing snow and ice interiors, and ionospheric characterization and correction in low-frequency SAR systems. His technological work covers SAR image formation, interferometry, multistatic synchronization, and MIMO processing, including demonstrations of high-resolution automotive SAR and coherent bistatic tomography using two aircraft. He is currently Scientific Coordinator of ESA's BIOMASS Data Innovation and Science Cluster (DISC) and Principal Investigator of the FISA URANUS project on multistatic UAV SAR formations. He has authored over 80 peer-reviewed journal papers, taught Remote Sensing and Signal Processing since 2013, and supervised numerous M.Sc. theses and Ph.D. students.
\end{IEEEbiography}

\end{document}